\documentclass[twocolumn,superscriptaddress,floatfix,preprintnumbers,amssymb ,amsmath]{revtex4}
\usepackage{graphicx}% Include figure files
\usepackage{dcolumn}% Align table columns on decimal point
\usepackage{bm}% bold math
\usepackage[latin1]{inputenc}
\usepackage[mathscr]{eucal}
\usepackage{epsfig}

\begin{document}
\title{Statistics of the dissipated energy in driven single-electron transitions}
\author{D.V. Averin}
\affiliation{Department of Physics and Astronomy, Stony Brook University, SUNY, Stony Brook, NY 11794-3800, USA}
\author{J.P. Pekola}
\affiliation{Low Temperature Laboratory, Aalto University, P.O. Box 13500, 00076 Aalto, Finland}

\begin{abstract}
We analyze the distribution of heat generated in driven single-electron transitions and discuss the related non-equilibrium work theorems. In the adiabatic limit, the heat distribution is shown to become Gaussian, with the heat noise that, in spite of thermal fluctuations, vanishes together with the average dissipated energy. We show that the transitions satisfy Jarzynski equality for arbitrary drive and calculate the probability of the negative heat values. We also derive a general condition on the heat distribution that generalizes the Bochkov-Kuzovlev equality and connects it to the Jarzynski equality.

\end{abstract}

%\date{\today}

\maketitle
Landauer principle \cite{lan,ben}, linking the erasure of information to the heat generated  in a computation process, plays an important role both in understanding the foundations of thermodynamics (see, e.g., \cite{rmp} and references therein) and in practical attempts to realize thermodynamically
reversible computation (see, e.g., \cite{nsq}). From physics point of view, a typical elementary step of a computation process is an  externally-driven and controlled switching between two distinguishable states of a thermodynamic system. From this perspective, one of the  basic facts of statistical mechanics underlying the Landauer principle is the statement that any system manipulated adiabatically, at frequencies below its energy relaxation rate, remains in a state close to the instantaneous thermal equilibrium. As a result, the energy is transferred into or out of the system reversibly, with the overall increase of the entropy ``of the universe'' that can be made arbitrarily small. Landauer principle addresses the average values, whereas the statistical fluctuations can lead to its violation in any given realization of the switching process \cite{mod1,mod2,mod3}. These fluctuations are particularly important in small systems, making it interesting to understand not only the behavior of the averages, but also the fluctuations. The question of the distribution of the generated heat, or of the work done on the system, is addressed in fluctuation theorems \cite{je,cr,hn} that have been of interest in the non-equilibrium statistical mechanics at least for the last 30 years \cite{bk}. These theorems, however, describe only some general characteristics of the heat/work distributions (discussed in more details below), while the distributions themselves are known only for some toy models (see, e.g., \cite{com}). The aim of this work is to calculate the distribution of the generated heat for one of the most basic models of nanoscale information processing: "single-electron tunneling" (SET) \cite{al}. There, the information is encoded in the position of individual electrons in a system of mesoscopic conductors, and is processed through electron tunneling between the neighboring conductors. For the purpose of studying the heat distribution, this model combines several attractive features. It is understood to a high precision, sufficient, e.g.,  for the development of metrological applications \cite{ns1,ns2}; it is considered as the basis of a practical scheme of reversible computing \cite{rev1,rev2,rev3}, and is sufficiently simple for explicit calculations. Our results show that, counter-intuitively, even for discrete  electron transitions, the width of the generated heat distribution  can be smaller than the energy $k_BT$ of thermal fluctuations, and  vanishes for adiabatic switching together with the average dissipated energy. We also discuss the applicability to the SET switching of the "Jarzynski equality" (JE) \cite{je}, one of the better-known work theorems, and derive an equation that generalizes a physically appealing but more limited equality derived earlier by Bochkov and Kuzovlev (BK) \cite{bk}.
\begin{figure}
    \includegraphics[width=8.5cm]{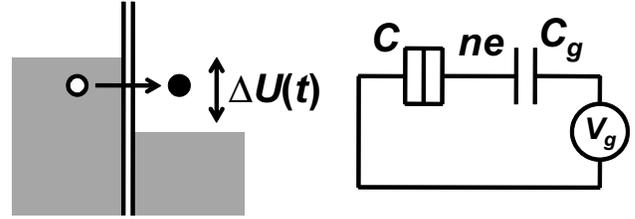}
    \caption{Driven single-electron transitions. Left panel: electrons jump between two electrodes across a tunnel barrier, with an external potential controlling the energy bias $\Delta U$ of the transition. If an electron tunnels with the bias $\Delta U$, this amount of energy is dissipated as heat.  Right panel: a basic realization of the transitions is provided by the single-electron box, where the gate voltage $V_g$ creates the bias via capacitance $C_g$. Electrons tunnel through the junction with capacitance $C$ and create charge $ne$ on the box island.}
    \label{fig1}
\end{figure}

Our discussion applies to driven single-electron transitions in general, although for the sake of clarity, we consider the specific example of a single-electron box (SEB) \cite{mb,sac}, a small conducting island coupled through a tunnel junction with conductance $G_T$ and capacitance $C$ to a large external electrode, and biased with respect to it by the gate voltage $V_g$ applied through the capacitance $C_G$, see Fig. \ref{fig1}. The island can carry extra electric charge $ne$. In the simplest regime of relatively low temperatures $T$, $\beta^{-1} \ll E_C \equiv e^2/(2C_\Sigma)$, where $\beta \equiv 1/k_B T$ and $C_\Sigma =C+C_g$ is the total capacitance of the box, and the gate voltage $V_g$ is restricted to the range $0 \le n_g \le 1$, where $n_g\equiv -C_gV_g/e$, the box dynamics can be limited to just two possible states, electron on the island, $n=1$, and electron off the island, $n=0$. We consider the basic switching dynamics between these two charge states with the ramp of the gate voltage across its value $\delta n_g \equiv n_g-1/2 =0$ at which these states are degenerate in energy. We first focus on the regime when the time dependence $\delta n_g (t)$ is not very rapid on the large scale set by the energy $E_C$, so that the transition $n=0\, \rightarrow n=1$ happens with certainty, while it can still be rapid on a smaller thermal scale $\beta^{-1}$. Such a transition is in general accompanied by dissipation of energy. The electrodes of the normal-metal SEB can be modeled as systems of non-interacting electrons. When an electron with energy $\epsilon$ tunnels into or out of such a system, it carries an amount of heat $\epsilon-\mu$, where $\mu$ is the electron chemical potential of the electrode. The energy $\epsilon$ is conserved in tunneling, and therefore the total amount of heat deposited in the two electrodes in electron transfer process is equal to the difference $\Delta U(t)$ between the chemical potentials of the electrodes at the time $t$ of the transition. In the SEB, $\Delta U(t)= 2E_C\delta n_g(t)$, and the total heat $Q$ generated in one ramp of the gate voltage is
\begin{equation} \label{eq1}
Q =  2E_C \sum_i \pm \delta n_g (\tau_i) = 2E_C \int dt \delta n_g (t) I(t) \, ,
\end{equation}
where the sum is over all back-and-forth ($\pm$) electron transitions between the states $n=0$ and $n=1$ at the time instants $\tau_i$ during one ramp, and $I(t)= \sum_i \pm \delta (\tau_i)$ is the associated particle current in the junction. The heat $Q$ is an experimental observable, which can in principle be determined either by direct measurement, or by detecting the position of the gate $n_g(\tau_i)$ at each transition. To calculate the distribution of heat $Q$ over an ensemble of transfer processes, we consider the situation of small junction conductance, $G_T\ll e^2/\hbar$, when the  dynamics of tunneling can be described by Markovian master equation for the occupation probabilities $p_n$ of the states $n$, assuming first the tunneling rates $\Gamma_\pm =\pm (G_T/e^2) \Delta U/(1-e^{\mp \beta \Delta U})$ characteristic for the normal-metal junction \cite{al}. The master equation for the occupation probabilities $p_n$, of the two charge states, $n=0,1$ reduces then to
$\dot{p}=\Gamma_D -\Gamma_{\Sigma}p$ for $p=p_1-p_0$,
where $\Gamma_D\equiv \Gamma_+ -\Gamma_-=G_T \Delta U/e^2$, and $\Gamma_{\Sigma} \equiv \Gamma_+ + \Gamma_-=(G_T \Delta U/e^2)\coth (\beta \Delta U /2)$. It gives the probability $p_{jk}(t_1,t_2)$ for the system to be in the state $k$ at time $t_2$, if it was in the state $j$ at time $t_1<t_2$ as
\begin{eqnarray}
p_{jk}(t_1,t_2) = \frac{1}{2} \Big[1+(-1)^{j+k} e^{ -
\int_{t_1}^{t_2} d \tau \Gamma_{\Sigma}(\tau) }
\nonumber \\
 - (-1)^k \int_{t_1}^{t_2} d \tau \Gamma_D(\tau) e^{ - \int_{\tau}^{t_2} d \tau'\Gamma_{\Sigma}(\tau') } \Big] \, .
\label{e1} \end{eqnarray}
To obtain the quantitative results for distribution of the generated heat $Q$, we take the time dependence of the gate voltage in Eq.~(\ref{eq1}) to be linear, $2E_C \delta n_g (t)=\eta t$, on the relevant small scale set roughly by $T$. (It can still be non-linear on a larger scale set by $E_C>>\beta^{-1}$.) Since the transitions, and hence the current $I(t)$, are suppressed away from the resonance, $t \rightarrow \pm \infty$, one can separate out the  average heat $\langle Q\rangle$ and integrate by parts in Eq.~(\ref{eq1}), reducing the equation for heat fluctuations $\tilde{Q} = Q-\langle Q\rangle$ in this case to $\tilde{Q}=\eta \int dt \tilde{n} (t)$, where $\tilde{n}(t)$ is the random realization of the charge on the SEB relative to its average in the gate ramp, $\dot{\tilde{n}}(t)= I(t)-\langle I(t) \rangle$. This means that the central moments of the heat fluctuations can be calculated through the correlation functions $K^{(m)}$ of the charge:
\begin{equation} \label{e2}
\langle \tilde{Q}^m \rangle =  \eta^m \int dt_1 ... dt_m K^{(m)} (t_1, ... ,t_m) \, ,
\end{equation}
which in turn can be obtained directly from the solution (\ref{e1}) of the master equation:
\begin{eqnarray}
K^{(m)} (t_1, ... ,t_m) = p_j(t_1)\tilde{n}_j(t_1) p_{jk}(t_1,t_2) \tilde{n}_k(t_2)  \nonumber \\ ... \,    p_{li}(t_{m-1},t_m)\tilde{n}_i(t_m) \, ,  \label{e3} \end{eqnarray}
where $p_j(t) \equiv p_{0j}(-\infty,t)$ are the probabilities of the two charge states evolving from the state $n=0$ at $t\rightarrow -\infty$, and $\tilde{n}_j(t)=j- \langle n(t) \rangle =j-p_1(t)$ is the value of the charge fluctuation $\tilde{n}(t)$ in the state $j=0,1$. Summation over all repeated indices is implied in Eq.~(\ref{e3}).

Equations (\ref{e1}) - (\ref{e3}) allow one, in principle, to find any central moment of the heat distribution.
We calculate explicitly the heat noise $\sigma_Q\equiv \langle \tilde{Q}^2  \rangle^{1/2}$ and the third moment $\lambda_Q \equiv \langle \tilde{Q}^3  \rangle^{1/3}$ which coincides with the third cumulant of the distribution. For $m=2$, one gets:
\begin{equation} \label{e4}
K^{(2)} (t_1,t_2) = p_0(t_1)p_1(t_1) e^{ - \int_{t_1}^{t_2} d \tau \Gamma_{\Sigma}(\tau) } , \;\;\; t_1\leq t_2\, .
\end{equation}
Since the charge dynamics we are considering is essentially classical, the correlator at $t_1>t_2$ is determined by the condition $K^{(2)} (t_1,t_2)=K^{(2)} (t_2,t_1)$. The correlator (\ref{e4}) substituted in Eq.~(\ref{e2}) gives the heat noise $\sigma_Q$.

\begin{figure}
\includegraphics[width=8.5cm]{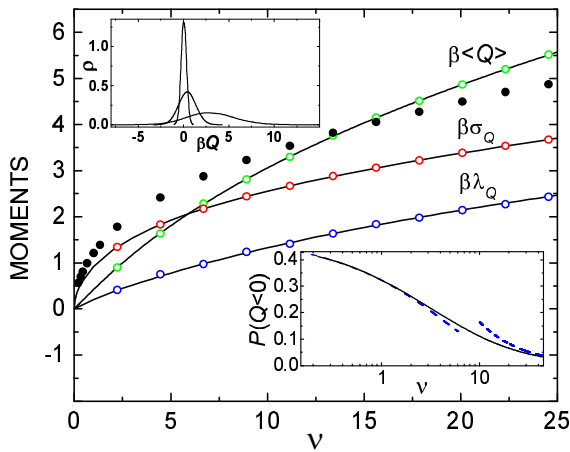}
\caption{The average and the higher moments of the distribution of the dissipated heat in single-electron transitions as functions of the ramp rate of the gate voltage driving the transitions. For comparison, open dots show the results (including the fourth central moment -- dots without a line) of direct Monte Carlo simulations of the transitions. The simulations also give the overall shape of the distribution (upper inset). Increasing ramp rate, $\nu =0.1, \, 1,\, 10$, leads to broadening of the distribution. Probability of extracting energy from thermal fluctuations in the transitions is shown in the lower inset. Solid curve is the result of the Monte Carlo simulations, with the two asymptotics (dashed curves) given by Eq.~(\ref{e9}).}
\label{f2} \end{figure}

The first few moments of the generated heat are shown in Fig.~\ref{f2} as functions of the dimensionless ramp rate $\nu \equiv \eta (e\beta)^2/G_T$. Despite the thermal fluctuations of energies  of tunneling electrons, the heat noise $\sigma_Q$ becomes smaller than $k_B T$ and vanishes together with the average dissipated energy $\langle Q \rangle$ in the limit of adiabatic reversible evolution, $\nu \rightarrow 0$. This means that, similarly to continuous diffusive processes \cite{com}, for discrete SET transitions, the dissipation is suppressed in the adiabatic limit not only on average, but for the individual ramps as well. Quantitatively, the instantaneous equilibrium probability $p = \Gamma_D/\Gamma_{\Sigma}$ obtained from Eq.~(\ref{e1}), combined with Eq.~(\ref{e2}) and (\ref{e4}) gives the heat noise for $\nu \le 1$:
\begin{equation} \label{e5}
\sigma_Q^2 = c \nu (k_BT)^2 = 2k_B T \langle Q\rangle ,
\end{equation}
with $c=(8/\pi^2) \sum_{n=0}^{\infty} (2n+1)^{-3} \simeq 0.85$.
The second equality in (\ref{e5}) follows from comparison to the  average dissipated energy $\langle Q\rangle = \eta \int dt t \langle \dot{n}(t) \rangle$ obtained using Eq.~(\ref{e1}) to find the first-order adiabatic correction to $\langle n(t)\rangle=p_1(t)$. This equality resembles the standard fluctuation-dissipation theorem relating (in this case, quasi-) equilibrium fluctuations of the dissipated energy and the average linear response to a slow change inducing this dissipation. Numerical evaluation of both quantities for arbitrary $\nu$ (as in Fig.~\ref{f2}) shows that this relation holds very accurately up to $\nu \simeq 5$. Used together with Eq.~(\ref{eq3}) below, it also means that the distribution function $\rho (Q)$ of the dissipated energy is Gaussian,
\begin{equation} \label{e6}
\rho (Q) = (\beta /4\pi \langle Q\rangle)^{1/2} \,  e^{-\beta (Q-\langle Q\rangle)^2/ 4 \langle Q\rangle} .
\end{equation}
This conclusion agrees with the behavior of $\lambda_Q$ shown in Fig.~\ref{f2}, which, as described below, satisfies $\lambda_Q \propto \nu \ll \sigma_Q \propto \sqrt{\nu}$ for $\nu \ll 1$. In the adiabatic limit, $Q$ is produced by many back-and-forth transitions and should indeed become Gaussian by the central limit theorem.

In the opposite limit of the rapid gate voltage ramp, $\nu \gg 1$, the heat distribution can be found in the  ``single-jump'' approximation assuming that in this case, the system has time for only one electron transition from the initial ($n=0$) to the final ($n=1$) charge state. Then,
\begin{equation} \label{e7}
\rho (x=\beta Q) = \frac{x/\nu}{1-e^{-x}} \exp\{-\frac{1}{\nu} \int_{-\infty}^{x} \frac{u}{1-e^{-u}}\} \, .
\end{equation}
One can check that this distribution reproduces well the large $\nu$ behavior of the moments plotted in Fig.~\ref{f2}. The leading part of these asymptotes can be obtained analytically by completely neglecting
the $1/\nu$ parts of the distribution, reducing it to $\rho(Q)= (Q/\sigma^2)  e^{- Q^2/2\sigma^2 }$, where $\sigma \equiv \sqrt{\nu}/
\beta = (\eta e^2/G_T)^{1/2} $, and giving, e.g., for the third cumulant $\lambda_Q= [(\pi-3)\sqrt{\pi/2}]^{1/3} \sigma$.

For arbitrary $\nu$, the third cumulant $\lambda_Q$ can be calculated from Eqs.~(\ref{e2}) and (\ref{e3}). Equation (\ref{e3}) combined with Eq.~(\ref{e1}) gives after some algebra for $m=3$:
\begin{eqnarray}
K^{(3)} (t_1,t_2,t_3) = p_0(t_1)p_1(t_1) [p_1(t_2)-p_0(t_2)] \nonumber \\ e^{ - \int_{t_1}^{t_3} d \tau \Gamma_{\Sigma}(\tau) } , \;\;\; t_1\leq t_2\leq t_3 \, . \label{e8} \end{eqnarray}
In all other time intervals, the correlator $K^{(3)}$ is defined, similarly to $K^{(2)}$, by the condition that it is symmetric with respect to permutations of the time arguments. The integrals of Eq.~(\ref{e8}) as required in Eq.~(\ref{e2}) can be evaluated numerically and give $\lambda_Q$ shown in Fig.~\ref{f2}. For  $\nu\gg 1$, this can be done analytically, with the result agreeing with that of the large $\nu$ approximation given above. In the adiabatic limit $\nu\ll 1$, one can use the quasiequilibrium expression for the probabilities, $p_{0,1}=(1\mp \Gamma_D/\Gamma_{\Sigma})/2$, and get  $\lambda_Q=c'\nu k_B T$, with $c'\simeq 0.22$.

One of the interesting features of the statistics of the dissipated energy in the SET transitions considered in this work is the existence of the $Q<0$  region, i.e. finite probability $P$ of an SET transition taking place with extraction of energy from thermal fluctuations rather than with energy dissipation. This probability could be the basis of
implementation of an SET version of the ``Maxwell's demon'' \cite{dem}: One can envision extracting electrical energy from thermal fluctuations in multi-junction SET  circuits, with the help of a measurement/feedback loop, similar to the ``information-to-energy conversion'' demonstrated recently by applying rotating electric field to a dimeric particle \cite{mod3}. The magnitude of this probability can be calculated for slow and rapid gate voltage ramps from Eqs.~(\ref{e6}) and (\ref{e7}):
\begin{equation}
P=\left\{ \begin{array}{ll} (1-\mbox{erf} [(c\nu/8)^{1/2}])/2, & \;\;\; \nu \ll 1, \\ \pi^2/(6\nu), &\;\;\; \nu \gg 1. \end{array} \right.
\label{e9} \end{equation}
In the adiabatic regime, the probability can be large, $P\rightarrow 1/2$, at the cost of the typical value $Q_e$ of the extracted energy being small on the scale of thermal energy, $Q_e \sim \sqrt{\nu} k_B T$. In the opposite limit, $\nu\gg 1$, $P$ is small, but $Q_e \sim k_B T$. For arbitrary $\nu$, $P$ can be calculated by direct Monte Carlo simulation of the transitions, see Fig.~\ref{f2} (lower inset), which shows good agreement with the asymptotic behavior (\ref{e9}) in the two limits.

The SET transitions represent a very convenient system for testing the general work theorems \cite{bk,je,cr,hn,ut} of non-equilibrium statistical mechanics. To derive them, we adopt a more general set of assumptions than used above. We do not assume any specific energy dependence of the transition rates $\Gamma_\pm$ besides the detailed balance, $\Gamma_+(\tau_i) = e^{\beta \Delta U(\tau_i)}\Gamma_-(\tau_i)$, and in principle, allow the transitions to transfer more than one electron across the barrier. The following discussion is thus applicable, with small adjustments, also to normal metal/superconductor (NIS) junctions, which in the box geometry, in addition to SET transitions, demonstrate individual Andreev processes \cite{us} transferring pairs of electrons. We also allow the gate voltage to have arbitrary time dependence, and the evolution of the SEB to start and end not in a definite charge state. Instead of transition probabilities (\ref{e1}), it is convenient now to represent Markovian evolution of the SEB in terms of all possible multi-transition paths between initial $-\mathcal T/2$ and final $+\mathcal T/2$ times, so that the average of any function $f(Q)$ of heat is
\begin{equation} \label{eq2}
\langle f(Q)\rangle =  \sum_{j=0}^\infty \sum_{n=0}^1 p_n\langle f(Q)\rangle_{n,j}.
\end{equation}
Here, as before, we restrict the dynamics to the two charge states, $p_n$ is the initial probability of the state $n$, and the averages $\langle f(Q)\rangle_{n,j}$ for trajectories starting at $n$ and making $j$ back-and-forth transitions are:
\begin{eqnarray} \label{eq2a}
&&\langle f(Q)\rangle_{n,j}=\int_{-\mathcal T/2}^{\mathcal T/2}d\tau_{j} \int_{-\mathcal T/2}^{\tau_{j}}d\tau_{j-1}...\int_{-\mathcal T/2}^{\tau_2}d\tau_1 \nonumber \\ &&\rho_n (\tau_1,\tau_2,...,\tau_{j})f\big(Q(\tau_1, ...,\tau_{j})\big).
\end{eqnarray}
The probability density $\rho_n$ can be expressed in terms of the probability $\eta_\pm(\tau_i,\tau_{i+1})= e^{ -\int_{\tau_i}^{ \tau_{i+1}} \Gamma_\pm(t')dt'}$ that the system does not make a $\pm$ transition in the time interval from $\tau_i$ to $\tau_{i+1}$. For odd values of $j$,
\begin{eqnarray*}
\rho_n (\tau_1, ...,\tau_{j})=\tilde\rho_\pm(-\mathcal T/2,\tau_1)\tilde\rho_\mp(\tau_1,\tau_2)\cdot  ... \cdot \tilde\rho_\pm(\tau_{j-1},\tau_j) \nonumber \\
\eta_\mp(\tau_j, \mathcal T/2), \;\;\;\; \tilde\rho_\pm (\tau_i,\tau_{i+1})\equiv \eta_\pm(\tau_i,\tau_{i+1})\Gamma_\pm (\tau_{i+1}),
\end{eqnarray*}
where the two sets of indices correspond to the two states $n$. For even $j$, the expression is constructed similarly.

The detailed balance for tunneling rates implies immediately that the expression of the average of $f(Q)=e^{-\beta Q}= \Pi_{l=1}^{j} e^{\mp \beta \Delta U(\tau_l)}$, taken according to Eqs.~\eqref{eq2} and \eqref{eq2a}, coincides with the transition probability $w$ between the charge states in the time-reversed evolution \cite{cr2} with the same voltage ramp, i.e.
\begin{eqnarray} \label{eq3}
\langle e^{-\beta Q}\rangle = \sum_{i,f} p_i \, w (f\rightarrow i) ,
\end{eqnarray}
where $i,f=n(\mp \mathcal T/2)$ are the initial and final charge states of the system. This equation gives a general version of the work equalities discussed previously \cite{bk,je,cr,hn}, in the form convenient for the SET  transitions. In particular, if the gate voltage ramp satisfies the condition $\delta n_g(-t)=-\delta n_g(t)$, as, e.g., does the linear ramp discussed above, the tunneling rate has the symmetry  $\Gamma_-(-t)=\Gamma_+(t)$. In this case, transition probabilities are symmetric, $w (f\rightarrow i)=w (i\rightarrow f)$, and Eq.~\eqref{eq3} gives $\langle e^{-\beta Q}\rangle = 1$, the relation used above to derive the distribution \eqref{e6}, and by setting $\beta Q=\Delta S$, it is seen to be identical to the central equation (9.1) of Ref.~\cite{bk} of Bochkov and Kuzovlev. We see that it is satisfied here for arbitrary distribution of the initial probabilities $p_i$ of the charge states. If the ramp is not antisymmetric relative to the degeneracy point at $t=0$, the right-hand-side of Eq.~\eqref{eq3} can be different from 1.

For equilibrium initial state, the probabilities $p_i$ can be expressed through the equilibrium probabilities $p_f$ defined for the gate voltage at the end of the evolution:  $p_i=p_f e^{\beta(E_f-E_i-\Delta F)}$, where $\Delta F$ is the difference of the free energy for the final and the initial gate voltage values in the ramp, and $E_{i,f}$ are the energies of the corresponding charge states given by the box charging Hamiltonian $H(n) =E_C n^2 -U_gn$, where $U_g \equiv eC_gV_g/C_\Sigma$. This means that the energy difference between the initial and final charge states can be written as
\begin{equation} \label{e11}
E_f-E_i=\Delta H=-\int (U_g-E_C) dn(t) -\int n dU_g(t),
\end{equation}
where the first integral coincides with the heat (\ref{eq1}), while the second term defines the "thermodynamic" work  \cite{com,j07} $W_{\rm th} = -\int n dU_g$. While this work is different from the actual physical work \cite{vil,chen} done on the box by the source of the gate voltage, which is given by $\int U_g dn$, it describes the change of the energies of the charge states.  The average  $\langle e^{-\beta W_{\rm th}}\rangle$ over the distribution of the thermodynamic work in a ramp can expressed similarly to Eq.~\eqref{eq3}. Relating then the probabilities $p_i$ through $p_f$ in this expression with the help of Eq.~(\ref{e11}), one gets the Jarzynski equality
\begin{eqnarray} \label{eqJE}
\langle e^{-\beta (W_{\rm th}-\Delta F)}\rangle = 1
\end{eqnarray}
for an arbitrary gate voltage ramp. Overheating effects were not considered in this analysis.

In summary, we obtained the distribution of heat generated in driven single-electron transitions and its central moments, and analyzed the most common work theorems and their validity in this context. The authors would like to acknowledge useful discussions with T. Ala-Nissil\"a, K.K. Likharev, M. M\"ott\"onen and O.-P. Saira.


\begin{thebibliography}{99}

\bibitem{lan} R. Landauer, IBM J.\ Res.\ Devel. {\bf 3}, 183 (1961).

\bibitem{ben} C. Bennett, IBM J.\ Res.\ Devel. {\bf 17}, 525 (1973).

\bibitem{rmp} K. Maruyama, F. Nori, and V. Vedral, Rev.\ Mod.\ Phys. {\bf 81}, 1 (2009).

\bibitem{nsq} J. Ren {\sl et al.},
IEEE Trans.\ Appl.\ Supercond. {\bf 19}, 961 (2009).

\bibitem{mod1} R. Dillenschneider and E. Lutz, Phys.\ Rev.\ Lett. {\bf 102}, 210601
(2009).

\bibitem{mod2} T. Sagawa and M. Ueda, Phys.\ Rev.\ Lett. {\bf 102}, 250602 (2009).

\bibitem{mod3} S. Toyabe {\sl et al.}, Nat.\ Phys. {\bf 6}, 988 (2010).

\bibitem{je} C. Jarzynski, Phys.\ Rev.\ Lett. {\bf 78}, 2690 (1997).

\bibitem{cr} G.E. Crooks, Phys.\ Rev. E {\bf 60}, 2721 (1999).

\bibitem{hn} M. Campisi, P. H\"anggi, and P. Talkner, arXiv:1012.2268.

\bibitem{bk} G.N. Bochkov and Yu.E. Kuzovlev, Physica A {\bf 106}, 443 (1981).

\bibitem{com} J. Horowitz and C. Jarzynski, Phys.\ Rev.\ Lett. {\bf 101}, 098901 (2008).

\bibitem{al} D.V. Averin and K. K. Likharev, in: {\em Mesoscopic Phenomena in Solids},
ed.\ by B.L. Altshuler, P.A. Lee, and R.A. Webb, (Elsevier: Amsterdam, 1991) p.\ 173.

\bibitem{ns1} J.P. Pekola {\sl et al.}, Nat. Phys. {\bf 4}, 120 (2008).

\bibitem{ns2} V.F. Maisi {\sl et al.}, New J. Phys. {\bf 11}, 113057 (2009).

\bibitem{rev1} K.K. Likharev and A.N. Korotkov, Science {\bf 273}, 763 (1996).

\bibitem{rev2} I. Amlani {\sl et al.}, Science {\bf 284}, 289 (1999).

\bibitem{rev3} A.O. Orlov {\sl et al.}, Appl.\ Phys.\ Lett. {\bf 77}, 295 (2000).

\bibitem{mb} M. B\"uttiker, Phys.\ Rev. B {\bf 36}, 3548 (1987).

\bibitem{sac} P. Lafarge {\sl et al.}, Z.\ Phys. B {\bf 85}, 327 (1991).

\bibitem{j07} C. Jarzynski, C.R. Physique {\bf 8}, 495 (2007).

\bibitem{dem} {\em Maxwell's demon}, ed.\ by H.S. Leff and A.F. Rex, (IoP: Bristol, 2003).

\bibitem{ut} Y. Utsumi {\sl et al.}, Phys. Rev. B {\bf 81}, 125331 (2010).

\bibitem{us} V.F. Maisi {\sl et al.}, arXiv:1012.5750.

\bibitem{cr2} G.E. Crooks, J.\ Stat.\ Phys. {\bf 90}, 1481 (1998).

\bibitem{vil} J.M.G. Vilar and J.M. Rubi, Phys.\ Rev.\ Lett. {\bf 100}, 020601 (2008).

\bibitem{chen} L.Y. Chen, J.\ Chem.\ Phys. {\bf 129}, 144113 (2008).

\end{thebibliography}
\end{document}